\documentclass[12pt]{article}

\usepackage{amsmath,amssymb,amsthm}
\usepackage{mathrsfs}
\usepackage{stmaryrd}
\usepackage{geometry}
\usepackage[hidelinks]{hyperref}
\usepackage{authblk}

\newcommand{\TBox}{\mathsf{T}}
\newcommand{\ABox}{\mathsf{A}}

\newcommand{\OBox}{\mathsf{O}}

\title{\bf TAPO-Structured Description Logic for Information Behavior: Procedural and Oracle-Based Extensions}
\author{\Large Takao Inou\'{e}}

\affil{\large Faculty of Informatics, Yamato University, \\ Osaka, Japan\footnote{Email: inoue.takao@yamato-u.ac.jp; \\ Personal Email: takaoapple@gmail.com \\ [I prefer my personal email address for correspondence.]}} 
\date{February 19,  2026}

\begin{document}
\maketitle

%
%
%

\begin{abstract}
We introduce \emph{TAPO-Structured Description Logic} (TAPO--DL), 
a formal extension of classical description logic designed to model 
\emph{information behavior} as a structured, dynamic process.

TAPO--DL extends the standard T--Box/A--Box architecture with two additional layers:
a \emph{Procedural Box} (P--Box), which supports concept-driven, imperative-style programs
such as conditional and iterative actions, and an \emph{Oracle Box} (O--Box),
which formalizes controlled interaction with external information sources.
While the terminological and assertional components capture static conceptual
and factual knowledge, the procedural and oracle-based components enable the
explicit representation of information-generating actions and external validation.

We provide a unified semantic framework for TAPO--DL based on a co-generative,
sheaf-theoretic interpretation, in which local informational states are modeled
as sections and informational stability corresponds to the existence of coherent
global structures. Within this setting, informational truth is characterized
as stability under repeated agentive interaction rather than correspondence
to a fixed global state.

By integrating description logic with procedural dynamics, oracle-based reasoning,
and sheaf-theoretic semantics, TAPO--DL offers a principled formal framework
for analyzing information behavior in contexts involving interaction,
uncertainty, and contextuality.
\\
\medskip

\noindent Keywords: Information Behavior,
Description Logic,
TAPO-Structured Description Logic,
Procedural Semantics,
Oracle-Based Reasoning,
Sheaf-Theoretic Semantics,
Knowledge Representation.
\\
\medskip

\noindent MSC2020: 03B45, 68T27, 18F20.
\end{abstract}

\tableofcontents

\section{Introduction}
Classical description logics (DLs) organize knowledge into a T--Box (concept
axioms) and an A--Box (individual assertions). This architecture is well--suited
for static knowledge representation, but it does not directly model key
phenomena of \emph{information behavior}: iterative search, conditional actions
based on partial information, and controlled interaction with external
resources.

This paper introduces TAPO-Structured Description Logic (for short, \emph{TAPO--DL}), extending the standard T/A--Box structure
with two additional components:
\begin{itemize}
  \item a \emph{P--Box}, which is \emph{not} a set of inference rules but a
        \emph{programmable} layer where one can write concept--driven procedures
        using imperative constructs (e.g., \textsf{if}--\textsf{then} and
        \textsf{while});
  \item an \emph{O--Box}, specifying admissible oracle interactions through which
        the system may incorporate externally obtained information.
\end{itemize}
The four components---Terminological, Assertional, Procedural, and Oracle-based---are
integrated via a sheaf--theoretic semantics of contextual information.

\section{Signature and Concept Language}
Let the signature be
\[
\Sigma = (N_C, N_R, N_I, \mathcal U),
\]
where $N_C$ is a set of concept names, $N_R$ a set of role names, $N_I$ a set of
individual names, and $\mathcal U$ a collection of contextual domains (opens,
situations, or information states).

Concept expressions are generated by the standard \textsf{ALC} grammar:
\[
C ::= \top \mid \bot \mid A \mid C \sqcap D \mid C \sqcup D \mid \neg C \mid
      \exists r.C \mid \forall r.C,
\]
with $A \in N_C$ and $r \in N_R$.

\section{T--Box and A--Box}
\subsection{T--Box (Concept Axioms)}
A T--Box $\TBox$ is a finite set of concept inclusions
\[
C \sqsubseteq D.
\]
These axioms are interpreted globally and are independent of contextual
annotations.

\subsection{A--Box (Contextual Assertions)}
An A--Box $\ABox$ consists of assertions of the form
\[
a : C@U \qquad (U \in \mathcal U),
\]
(and role assertions $(a,b):r@U$). The annotation $@U$ indicates that the
assertion holds \emph{with respect to the contextual domain $U$}. Intuitively,
$C@U$ denotes the localization of the concept $C$ to the information available
in $U$.

\section{Context as Sheaf Semantics}
We interpret contexts as objects of a site (or, for simplicity, as opens of a
topological space). Concepts are interpreted as sheaves over the partially
ordered set $(\mathcal U, \subseteq)$.

For each concept $C$ and context $U$, the interpretation $C(U)$ denotes the set
of individuals satisfying $C$ under the information available in $U$. If
$V \subseteq U$, there is a restriction map
\[
\rho_{UV}: C(U) \to C(V),
\]
ensuring contextual monotonicity:
\[
a:C@U \Rightarrow a:C@V \quad (V \subseteq U).
\]

\section{P--Box: A Minimal Imperative Language}
\subsection{Knowledge States}
A knowledge state is a pair
\[
\Sigma = (\TBox, \ABox),
\]
where $\TBox$ is a fixed T--Box and $\ABox$ is the current A--Box.

\subsection{Guards}
Guards (conditions) are given by
\[
\varphi ::= \top \mid \bot \mid \alpha \mid \neg\varphi \mid (\varphi \wedge \varphi),
\]
where atomic guards are
\[
\alpha ::= a:C@U \mid (a,b):r@U \mid (C \sqsubseteq D).
\]
The satisfaction relation $\Sigma \models \varphi$ is defined as follows:
\begin{itemize}
  \item $(\TBox,\ABox)\models a:C@U$ iff $a:C@U\in\ABox$;
  \item $(\TBox,\ABox)\models (a,b):r@U$ iff $(a,b):r@U\in\ABox$;
  \item $(\TBox,\ABox)\models (C\sqsubseteq D)$ iff $\TBox\vdash_{\mathrm{DL}} C\sqsubseteq D$;
\end{itemize}
with Boolean connectives interpreted classically.

\subsection{Program Syntax}
The P--Box language $\mathbf P_0$ consists of programs generated by
\[
P ::= \mathbf{skip}
 \mid \mathbf{add}\;\beta
 \mid \mathbf{del}\;\beta
 \mid P;P
 \mid \mathbf{if}\;\varphi\;\mathbf{then}\;P\;\mathbf{else}\;P
 \mid \mathbf{while}\;\varphi\;\mathbf{do}\;P,
\]
where $\beta$ ranges over A--Box assertions (either $a:C@U$ or $(a,b):r@U$).

\subsection{Operational Semantics (Big-step)}
We write
\[
\langle P, \Sigma \rangle \Downarrow \Sigma'
\]
for the big--step evaluation relation. The core rules are standard.

\paragraph{Basic commands.}
\begin{align*}
\langle \mathbf{skip},(\TBox,\ABox)\rangle &\Downarrow (\TBox,\ABox),\\
\langle \mathbf{add}\;\beta,(\TBox,\ABox)\rangle &\Downarrow (\TBox,\ABox\cup\{\beta\}),\\
\langle \mathbf{del}\;\beta,(\TBox,\ABox)\rangle &\Downarrow (\TBox,\ABox\setminus\{\beta\}).
\end{align*}

\paragraph{Sequencing.}
\[
\frac{\langle P_1,\Sigma\rangle\Downarrow\Sigma_1\quad \langle P_2,\Sigma_1\rangle\Downarrow\Sigma_2}
{\langle P_1;P_2,\Sigma\rangle\Downarrow\Sigma_2}
\]

\paragraph{Conditionals.}
\[
\frac{\Sigma\models\varphi\quad \langle P_1,\Sigma\rangle\Downarrow\Sigma'}
{\langle \mathbf{if}\;\varphi\;\mathbf{then}\;P_1\;\mathbf{else}\;P_2,\Sigma\rangle\Downarrow\Sigma'}
\qquad
\frac{\Sigma\not\models\varphi\quad \langle P_2,\Sigma\rangle\Downarrow\Sigma'}
{\langle \mathbf{if}\;\varphi\;\mathbf{then}\;P_1\;\mathbf{else}\;P_2,\Sigma\rangle\Downarrow\Sigma'}
\]

The intended meaning of the first inference rule (left rule) is as follows.

\begin{itemize}
  \item The current state $\Sigma$ satisfies the condition $\varphi$, that is,
  $\Sigma \models \varphi$.
  \item Executing the procedure $P_1$ from the state $\Sigma$
  evaluates to the resulting state $\Sigma'$, written as
  $\langle P_1, \Sigma \rangle \Downarrow \Sigma'$.
\end{itemize}

Under these assumptions, the execution of the conditional statement
\[
\mathbf{if}\;\varphi\;\mathbf{then}\;P_1\;\mathbf{else}\;P_2
\]
from the state $\Sigma$ also evaluates to the state $\Sigma'$.
In this case, the \texttt{else} branch $P_2$ is not executed.

The intended meaning of the second inference rule (right rule) is as follows.

\begin{itemize}
  \item The current state $\Sigma$ does not satisfy the condition $\varphi$, that is,
  $\Sigma \not\models \varphi$.
  \item Executing the procedure $P_2$ from the state $\Sigma$
  evaluates to the resulting state $\Sigma'$, written as
  $\langle P_2, \Sigma \rangle \Downarrow \Sigma'$.
\end{itemize}

Under these assumptions, the execution of the conditional statement
\[
\mathbf{if}\;\varphi\;\mathbf{then}\;P_1\;\mathbf{else}\;P_2
\]
from the state $\Sigma$ evaluates to the state $\Sigma'$.
In this case, the \texttt{then} branch $P_1$ is not executed.

\paragraph{While.}
\[
\frac{\Sigma\not\models\varphi}
{\langle \mathbf{while}\;\varphi\;\mathbf{do}\;P,\Sigma\rangle\Downarrow\Sigma}
\qquad
\frac{\Sigma\models\varphi\quad \langle P,\Sigma\rangle\Downarrow\Sigma_1\quad
\langle \mathbf{while}\;\varphi\;\mathbf{do}\;P,\Sigma_1\rangle\Downarrow\Sigma_2}
{\langle \mathbf{while}\;\varphi\;\mathbf{do}\;P,\Sigma\rangle\Downarrow\Sigma_2}
\]

The intended meaning of the first inference rule for the \texttt{while} construct is as follows.

\begin{itemize}
  \item The current state $\Sigma$ does not satisfy the condition $\varphi$, that is,
  $\Sigma \not\models \varphi$.
\end{itemize}

Under this assumption, the execution of the loop
\[
\mathbf{while}\;\varphi\;\mathbf{do}\;P
\]
from the state $\Sigma$ terminates immediately
and evaluates to the same state $\Sigma$.
In this case, the loop body $P$ is not executed.

The intended meaning of the second inference rule for the \texttt{while} construct is as follows.

\begin{itemize}
  \item The current state $\Sigma$ satisfies the condition $\varphi$, that is,
  $\Sigma \models \varphi$.
  \item Executing the loop body $P$ from the state $\Sigma$
  evaluates to an intermediate state $\Sigma_1$, written as
  $\langle P, \Sigma \rangle \Downarrow \Sigma_1$.
  \item Executing the loop again from the updated state $\Sigma_1$
  evaluates to the final state $\Sigma'$, written as
  $\langle \mathbf{while}\;\varphi\;\mathbf{do}\;P, \Sigma_1 \rangle \Downarrow \Sigma'$.
\end{itemize}

Under these assumptions, the execution of the loop
\[
\mathbf{while}\;\varphi\;\mathbf{do}\;P
\]
from the initial state $\Sigma$ evaluates to the state $\Sigma'$.

The \textsf{while} construct is therefore generally partial (non--termination is
possible), matching the open--ended nature of iterative information behavior.

\section{O--Box: Oracle Interaction Rules}
The O--Box specifies admissible interactions with external information sources
(oracles). An oracle call may introduce new contextual assertions into the
A--Box, representing openness of the system to the environment.

Formally, an O--Box can be treated as a relation
\[
\llbracket \OBox \rrbracket \subseteq (\TBox\times\ABox)\times(\TBox\times\ABox),
\]
whose transitions are externally justified (e.g., by an API response, a human
judgment, or a sensor reading), rather than internally derivable.

\section{An Application to Information Behavior}

In this section, we present a simple application of the co-generative ontological framework
to the analysis of information behavior.
Here, \emph{information behavior} refers to the structured actions by which an epistemic agent
selects, interprets, and stabilizes informational entities from a latent informational domain.

\subsection{Ontological Interpretation of Information}

Within the present axiomatization, informational entities are not assumed to exist
as fully determined objects \emph{a priori}.
Instead, they are treated as elements of the domain of structural potential,
which become manifest through interaction with epistemic agents.

Let $\mathcal{P}$ denote the domain of structural potential,
$\mathcal{E}$ the class of epistemic agents,
and $\mathcal{M}$ the domain of manifested existents, as introduced in previous sections.
An informational object $i$ initially belongs to $\mathcal{P}$ as a latent structure,
such as an uninterpreted signal, data stream, or symbolic pattern.

\subsection{Information Behavior as a Co-Generative Process}

An information behavior is modeled as a co-generative interaction
\[
(e, i) \longmapsto m,
\]
where $e \in \mathcal{E}$ is an epistemic agent,
$i \in \mathcal{P}$ is a latent informational structure,
and $m \in \mathcal{M}$ is a manifested informational object.

This process does not merely reveal pre-existing information,
but actively constitutes the informational object as meaningful for the agent.
Different agents, or different internal states of the same agent,
may generate distinct manifested informational objects from the same latent structure.
\medskip

\noindent \bf Remark\rm . This interpretation aligns with the view that information is not ontologically primitive, but emerges through structured interaction. The same data may remain informationally nonexistent for one agent, while becoming a stable informational entity for another.
\medskip

For the author's co-generative theory of existece, refer to Inou\'{e} \cite{Inoue2026Cog}.

\subsection{Stability and Informational Truth}

According to the co-generative axioms,
truth is interpreted as a stability phenomenon.
In the context of information behavior,
an informational object $m \in \mathcal{M}$ is said to be \emph{informationally stable}
if repeated interactions between $e$ and $i$ consistently regenerate $m$.

Such stability corresponds to the agent's recognition of the information
as reliable, meaningful, or true.
This allows informational truth to be agent-relative
without collapsing into arbitrariness,
since stability is constrained by the structure of $\mathcal{P}$.

\subsection{Minimal Example}

As a minimal example, consider a sensor system acting as an epistemic agent.
Raw sensor signals exist initially as latent structures in $\mathcal{P}$.
Through interpretive protocols implemented by the agent,
certain signal patterns are repeatedly stabilized as specific informational objects,
such as ``obstacle detected'' or ``temperature exceeds threshold''.

Within the present framework,
these informational objects exist precisely insofar as
they are co-generatively stabilized through agent–structure interaction.

\section{A Sheaf-Theoretic Interpretation of Information Behavior}

In this section, we strengthen the interpretation of information behavior
by making explicit use of sheaf-theoretic structures.
The central claim is that information behavior is naturally modeled
as the generation and stabilization of local sections,
together with their coherent gluing into global informational entities.

\subsection{Informational Domains as Sites}

Let $(\mathcal{C}, J)$ be a site,
where objects of $\mathcal{C}$ represent informational contexts
(e.g.\ temporal windows, sensor modalities, linguistic perspectives),
and $J$ is a Grothendieck topology encoding admissible coverings.

We interpret informational potential as a presheaf
\[
\mathcal{I} : \mathcal{C}^{\mathrm{op}} \to \mathbf{Set},
\]
where $\mathcal{I}(U)$ is the set of latent informational structures
available in context $U$.

\subsection{Epistemic Agents as Section-Generating Structures}

An epistemic agent $e \in \mathcal{E}$
is modeled as a structure that assigns,
to each context $U \in \mathcal{C}$,
a (possibly partial) selection of sections
\[
s_U \in \mathcal{I}(U),
\]
subject to interpretive constraints internal to the agent.

Information behavior is thus the process by which an agent
extracts, refines, or stabilizes local sections
from the presheaf $\mathcal{I}$.

\subsection{Sheaf Condition and Informational Coherence}

The sheaf condition plays a central ontological role.
Given a covering $\{U_i \to U\} \in J$,
a family of local informational sections
$\{s_{U_i} \in \mathcal{I}(U_i)\}$
represents distributed informational behavior.

If these sections agree on overlaps,
they admit a unique gluing
\[
s_U \in \mathcal{I}(U),
\]
which we interpret as the emergence of a coherent informational object
at the global level.

This gluing operation corresponds precisely to the transition
from fragmented, context-dependent information
to stabilized, meaningful informational existence.

\subsection{Manifestation as Stabilized Global Sections}

Within the co-generative ontology,
manifested informational entities correspond to
globally stabilized sections of $\mathcal{I}$.
That is, an informational object exists in $\mathcal{M}$
if and only if it arises as a glued section
that remains stable under further contextual refinements.
\medskip

\noindent Remark\rm . 
This interpretation explains why informational existence is neither purely subjective
nor purely objective.
Local sections depend on agents and contexts,
while global coherence is constrained by the topology $J$.

\subsection{Minimal Example: Distributed Sensing}

Consider a distributed sensor network.
Each sensor modality or time slice defines a context $U \in \mathcal{C}$,
with raw sensor readings forming local sections in $\mathcal{I}(U)$.

Information behavior consists in extracting these local sections
and verifying their compatibility across overlaps.
When coherence is achieved,
a global section emerges,
corresponding to an informational object such as
``an obstacle is present.''

In this framework,
informational truth is identified with the stability
of glued sections under refinement,
rather than with correspondence to a pre-given global state.

\section{Conclusion and the Ruture Work}
TAPO--DL extends description logic beyond static representation by introducing a
programmable procedural layer (P--Box) and explicit oracle interaction (O--Box).

In the future, we hope to develop TAPO-DL in a more sheaf-theoretic manner (cf. Inou\'{e} \cite{InoueDLLecture}).

$$ $$

\noindent Takao Inou\'{e}

\noindent Faculty of Informatics

\noindent Yamato University

\noindent Katayama-cho 2-5-1, Suita, Osaka, 564-0082, Japan

\noindent inoue.takao@yamato-u.ac.jp
 
\noindent (Personal) takaoapple@gmail.com (I prefer my personal mail)

\end{document}